# Chromium Supplementation And The Essentiality Of Chromium To Human Nutrition: A Narrative Review

Matthew Tirona

## ABSTRACT

This narrative review evaluates the effect of chromium supplementation on glycemia and serum lipids, with an emphasis on patients with type 2 diabetes mellitus (T2DM). Additionally, this narrative review evaluates the essentiality of the trace mineral chromium to human nutrition. Meta-analyses and reviews were included, while certain clinical trials were specifically included to discuss flaws or impact. Overall, this narrative review concludes that chromium supplementation likely has no beneficial effect on glycemia or serum lipids (in subjects with or without T2DM). This narrative review also concludes the essentiality of chromium to human nutrition has become increasingly challenged over time, with some investigators postulating that chromium is pharmacologically active rather than an essential trace mineral. However, further randomized controlled trials (RCTs) are necessary to come to solid conclusions about the effect of chromium supplementation and the essentiality of chromium to human nutrition. This investigation is necessary as manufacturers continue to market chromium supplements as helpful aids to T2DM patients based on flawed studies.

## INTRODUCTION

According to the International Diabetes Federation, approximately 537 million adults are living with diabetes worldwide [1]. This number is projected to increase to 643 million and 783 million by 2030 and 2045, respectively. Chromium has historically been deemed as an essential trace mineral to the metabolism of glucose, among other functions related to type 2 diabetes mellitus (T2DM) [6][12][15][16]. Thus, chromium supplementation has been often postulated to improve outcomes for diabetic patients. Manufacturers have widely advertised chromium supplements to people in the United States with T2DM, despite the FDA stating "the existence of such a relationship between chromium picolinate and either insulin resistance or type 2 diabetes is highly uncertain" [2][3].

In 2001, chromium was considered to be an essential nutrient by the Food and Nutrition Board (FNB) of the National Academies of Sciences, Engineering, and Medicine based on its effects on insulin action [2]. Recent research has however challenged this, hypothesizing that chromium might be beneficial in pharmacological amounts, rather than being an essential mineral [19]. Although the FNB has not reevaluated the essentiality of chromium since 2001, in 2014, the European Food Safety Authority (EFSA) Panel on Dietetic Products, Nutrition and Allergies concluded that "there is no evidence of beneficial effects associated with chromium intake in healthy subjects" nor should the Adequate Intake (AI) for chromium be set [4]. Nevertheless, AIs exist for chromium, set by the FNB in 2001 (Table 1), as the element is technically still considered essential in the United States and Canada [5].

**Table 1** Adequate Intakes (AIs) for Chromium

| Age | Male | Female |
|---|---|---|
| Birth–6 months* | 0.2 mcg | 0.2 mcg |
| 7–12 months* | 5.5 mcg | 5.5 mcg |
| 1–3 years | 11 mcg | 11 mcg |
| 4–8 years | 15 mcg | 15 mcg |
| 9–13 years | 25 mcg | 21 mcg |
| 14–18 years | 35 mcg | 24 mcg |
| 19–50 years | 35 mcg | 25 mcg |
| 51+ years | 30 mcg | 20 mcg |
| **Age** | **Pregnancy** | **Lactation** |
| 14–18 years | 29 mcg | 44 mcg |
| 19–50 years | 30 mcg | 45 mcg |

*For infants from birth to age 12 months, the AIs are based on the mean chromium intakes of infants fed primarily human milk and, for older infants, complementary foods* [2].

Even though the EFSA concluded that chromium intake has no evidence of being beneficial in healthy patients, there exists a debate on the effects of chromium, particularly chromium supplementation in high pharmacological doses for patients with T2DM. But, the efficacy of chromium supplementation as a means of improving health outcomes remains inconclusive. Through an analysis of relevant literature, this narrative review appraises the evidence and examines the debate surrounding the essentiality of chromium to human nutrition and the efficacy of chromium as a supplement, with an emphasis on patients with T2DM.

## GLYCEMIA

Numerous trials have evaluated the effect of supplemental chromium on glycemia. One 1997 clinical trial by Anderson et al hypothesized that the elevated intake of supplemental chromium is involved in the control of T2DM [6]. 180 men and women being treated for T2DM were divided randomly into three groups, the first being supplemented with placebo, the second with 100 mcg of chromium picolinate twice daily, and the third with 500 mcg of chromium picolinate twice daily. Subjects were instructed to continue to take normal medications and not change their eating and living habits. In the 1,000 mcg group (500 mcg two times a day), $HbA_{1c}$ values improved significantly after 2 months, while in the 200 mcg group (100 mcg two times a day), values improved moderately (placebo: 8.5 ± 0.2%, 200 mcg: 7.5 ± 0.2%, 1,000 mcg: 6.6 ± 0.1%). Fasting blood glucose (FBG) concentrations were also significantly lower in the group receiving 1,000 mcg after both 2 months and 4 months, but fasting glucose concentrations also decreased in the placebo group, albeit to a smaller degree.

The results of Anderson et al have prompted manufacturers to extensively advertise chromium as beneficial to the general public [7]. Although often cited by manufacturers, numerous meta-analyses have excluded Anderson et al or deemed it of poor quality [7][8][9][10]. A 2002 meta-analysis by Althuis et al noted the noncomparability of the population studied (China) and how its application to the Western hemisphere is uncertain. Althuis et al furthermore noted the potential poor nutritional status of the subjects in Anderson et al, due to their body mass index (BMI) of 22-23. Althuis et al ultimately stated that the data from randomized controlled trials (RCTs) showed no effect of chromium on glucose or insulin concentrations in nondiabetic subjects, while in subjects with T2DM, the results were inconclusive (in part due to Anderson et al).

A 2007 systematic review by Balk et al was particularly critical of Anderson et al, finding that it–along with almost half of the studies that met criteria–were of poor quality [8]. Balk et al also noted that Anderson et al was also the only study that showed statistically significant improvements in 2-h (120 minutes) postload glycemia with either 200 or 1,000 mcg/day chromium picolinate (the decrease was greater with the higher dose). Of course, the poor quality of Anderson et al undermines this, despite largely prompting further studies examining the dose effect of chromium picolinate. Balk et al highlighted how Anderson et al prompted studies that utilized higher (pharmacological) doses of chromium picolinate (400 or 1,000 mcg/day). Ultimately, Balk et al concluded that chromium had no significant effect on glucose (or lipid) metabolism in people without diabetes, while chromium supplementation modestly improved glycemia among patients with T2DM. However, Balk et al qualified that the overall poor quality limited the strength of the conclusion, and consequently, future RCTs are necessary. Additionally, a 2014 meta-analysis by Bailey ranked Anderson et al as lowest in quality–despite having a large sample size–due to the insufficient data provided needed to calculate preintervention standard deviations and standard errors and preintervention and postintervention means [9]. Thus, Bailey excluded Anderson et al in the meta-analysis, coming to the conclusion that chromium supplementation appears to provide no benefits to populations where chromium deficiency is unlikely.

A 2006 meta-analysis by Trumbo and Ellwood–who are with the Division of Nutrition Programs and Labeling of the US Food and Drug Administration (FDA) in College Park, Maryland–concluded that there is very little credible evidence that supports a qualified health claim for chromium picolinate and reduced risk of insulin resistance (and thus reduced risk of T2DM) and that such relationship is highly uncertain [11]. In 2005, the FDA stated:

> *"One small study suggests that chromium picolinate may reduce the risk of insulin resistance, and therefore possibly may reduce the risk of type 2 diabetes. FDA concludes, however, that the existence of such a relationship between chromium picolinate and either insulin resistance or type 2 diabetes is highly uncertain."* [3]

Although a 2013 meta-analysis by Abdollahi et al concluded that chromium lowers FBG (but does not affect $HbA_{1c}$, lipids, and BMI), Bailey noted how that meta-analysis failed to include studies that may have not supported an effect of chromium on FBG in diabetic subjects [9][12]. Bailey also noted Abdollahi et al eliminated a study based on its perceived quality, limiting the conclusions of Abdollahi et al which support a beneficial effect of chromium. A 2022 meta-analysis by Zhao et al concluded that the use of chromium supplementation can reduce the glycosylated hemoglobin of subjects with T2DM to an extent that carries statistical significance but little clinical significance [10]. Additionally, Zhao et al concluded that chromium supplementation cannot effectively improve the blood glucose (or lipid) levels of subjects with T2DM, although further RCTs of different types of chromium preparations are necessary for more accurate and higher quality results.

Some meta-analyses concluded that chromium supplementation could have beneficial effects, although such conclusions are limited. A 2014 meta-analysis by Suksomboon et al suggested that chromium supplementation has favorable effects on glycaemic control in patients with diabetes (and thus diabetics with inadequate glycemic control could benefit from supplementation), although the results are limited due to the high heterogeneity and small sample size [13]. A 2020 meta-analysis by Asbaghi et al concluded that chromium supplementation might improve glycemic control indices in T2DM patients, however, its findings are limited due to the flawed RCTs included and widely varying dosages, physiological status, and age groups [14]. Ultimately, the consensus is that chromium supplementation likely has no beneficial effect on glycemia and thus no beneficial effect for patients with T2DM. Nevertheless, further RCTs and studies are recommended to come to a solid conclusion.

## SERUM LIPIDS

Many studies of chromium supplementation on glycemia included reports of serum lipids as well. A 1995 clinical trial by Wilson and Gondy randomized 26 non-obese, non-diabetic adults (mean age of 36 years) into a chromium group or a placebo group, with the chromium group receiving 220 mcg of chromium daily in the form of gelatin capsules containing chromium(III) nicotinate [15]. It was concluded that chromium supplementation has no apparent favorable changes in serum lipids in those individuals. A 2000 clinical trial by Amato et al concluded similar results; 19 subjects (9 men and 10 women, aged 63-77) were either given 1,000 mcg of chromium picolinate daily or a placebo for 8 weeks [16]. No significant change in serum lipids was observed. However, both investigations are limited due to the relatively small number of subjects in each respective group.

The aforementioned meta-analysis by Balk et al concluded that chromium supplementation has no significant effect on lipid metabolism in people without diabetes, citing both of the previously mentioned studies [8]. Although a flawed meta-analysis, Abdollahi et al concluded chromium does not affect total cholesterol (TC), high-density lipoprotein concentration (HDL-C), low-density lipoprotein concentration (LDL-C), very-low-density lipoprotein concentration (VLDL-C), and triglycerides (TG) [12]. Zhao et al concluded the use of chromium supplements cannot effectively improve the serum lipid levels of patients with T2DM [10].

However, Suksomboon et al noted that chromium monosupplement may improve TG and HDL-C levels in diabetics [13]. Again, the conclusions of Suksomboon et al are limited due to the high heterogeneity and small sample size. A 2021 meta-analysis by Tarrahi et al showed that chromium supplementation has a beneficial effect on TC, while subgroup analysis showed a significant lowering effect of chromium supplementation on TC, TG, and VLDL [17]. Tarrahi et al also claim that chromium supplementation may be effective in diabetic patients with short-duration and low-dose. Nevertheless, Tarrahi et al assure that in the future, well-designed RCTs with a large population are necessary to clarify such claims. A recent 2022 umbrella review by Vajdi et al (which cites Suksomboon et al and Tarrahi et al) concludes that the available evidence proposes no beneficial effect of chromium supplementation on serum lipids in adults [18]. Note that Mohammad Javad Tarrahi is listed as an author for both Tarrahi et al and Vadji et al. Overall, it seems that chromium supplementation likely has no significant effect on serum lipids in patients with or without T2DM.

## ESSENTIALITY

For an element to be essential, three things must be true: (1) it has a defined biochemical function; (2) its lack causes death or reproduction failure; and (3) the addition of it to the diet can prevent such effects [19]. Over time, the essentiality of chromium has been increasingly challenged. In 1997, a review by Anderson concluded that chromium is an essential nutrient for sugar and fat metabolism, yet the dietary intake in the United States and most other developed countries is suboptimal, leading to widespread symptoms that are similar to those of diabetes and/or cardiovascular diseases (CVD) [20]. In 2001, a review by Krejpcio suggested that–although the exact function of chromium is not fully understood–chromium interacts with thyroid metabolism, with nucleic acids, and with insulin and its receptors [21]. Krejpcio assured that it was impossible to set definite recommendations for chromium supplementation for the general Polish population due to insufficient data.

In 2007, a review by Pechova and Pavlata noted the divided experiments that show chromium supplementation having a positive effect or no positive effect [22]. Also in 2007, in Chapter 3 of *The Nutritional Biochemistry of Chromium(III)*, "Multiple Hypotheses for chromium(III) biochemistry: Why the essentiality of chromium(III) is still questioned," Stearns concluded that the existence of glucose tolerance factor (Cr-GTF) had become less likely and that the role of chromium(III) in iron metabolism had not been adequately investigated to make a conclusion [23]. Stearns also concluded that the hormesis characteristic of chromium(III) should be explored as a possible solution to its apparent biological activity.

In 2017, a review by Vincent concluded that chromium(III) should be indicated as pharmacologically active and not as essential [19]. Whether consuming a nutrition-designed diet or a self-selected diet, an American consumes around 30 mcg of chromium daily [24][25]. Given that the AI for chromium is 35 mcg/day for adult men and 25 mcg/day for adult women, the AI suggests that over 98% of the American population receiving this quantity display no deficiency-caused health problems. Vincent cites a 2011 study by Di Bona et al as one that has "unambiguously demonstrated that chromium has a pharmacological rather than a nutritional effect." [26]. Note that Vincent is listed as a co-author for Di Bona et al. Ultimately, further investigation is necessary to solidify chromium as a pharmacologically active and not essential. However, it seems that an increasing amount of research undermines chromium as an essential nutrient for humans.

## CONCLUSION

Overall, it seems that chromium supplementation likely has no significant effect on glycemia and serum lipids. Of course, further studies are necessary to come to a solid conclusion on the effect of supplemental chromium. These studies need to ensure that patients are free of nutrition deficiencies to avoid misleading evidence. Additionally, the essentiality of chromium needs to be reconsidered worldwide to prevent the spread of misinformation by manufacturers targeting patients with T2DM. Particularly, the FNB needs to reevaluate the essentiality of chromium and take a clear stance.